\documentclass[aps,prd,showpacs]{revtex4}
\usepackage{graphicx}
\usepackage{epsfig}
\usepackage{psfig}
\usepackage{revsymb}
\begin{document}
\renewcommand{\thesection}{\arabic{section}}
\renewcommand{\thesubsection}{\arabic{subsection}}
\title{A Relativistic Generalization of Fowler-Nordheim Cold Emission in
Presence of Strong Magnetic Field}
\author{Arpita Ghosh and Somenath Chakrabarty$^\dagger$}
\affiliation{
Department of Physics, Visva-Bharati, Santiniketan 731 235, 
West Bengal, India\\ 
$^\ddagger$E-mail:somenath.chakrabarty@visva-bharati.ac.in}
\pacs{97.60.Jd, 97.60.-s, 75.25.+z} 
\begin{abstract}
A relativistic version of cold emission of electrons in presence of
strong magnetic field, relevant for strongly magnetized neutron stars is
obtained. It is found that in this scenario, a scalar type 
potential barrier does not allow quantum tunneling through the surface.
Whereas, in presence of a vector type surface barrier, the probability
of electron emission is much larger compared to the original
Fowler-Nordheim cold emission of electrons. It is found that the relativistic 
version in presence of strong magnetic field does not follow exponential decay.
\end{abstract}
\maketitle
\section{introduction}
There are mainly three kinds of electron emission
processes from metal surface, they are: (i) thermionic emission, (ii)
photoelectric emission and (iii) cold emission or field emission.

Thermionic emission is the external thermal energy induced flow of electrons 
through a metallic surface barrier into vacuum. This occurs 
because the thermal energy given to the electrons overcome the
confining barrier potential. Therefore, to have a thermo-electric current, 
the thermal energy or equivalently the temperature should have a lower critical value. 
This lower limit strongly depends on the work function of the material, 
which is the minimum energy needed to remove an electron with zero 
kinetic energy from the material surface. 

The photoelectric effect is a phenomenon in which electrons are emitted from 
matter as a result of absorption of electromagnetic energy in the form of quanta 
of the radiation incident on their surfaces. 
Analogous to the thermionic emission, in the case 
of photo-emission, the incident electromagnetic waves or photons must have a
lower critical frequency, below which there will be no photo-emission.
This lower limit also strongly depends on the work function of the material.

Finally the field emission (also known as cold emission) is an electron 
emission process induced by strong external electric fields at zero or 
extremely low temperature. Field emission can happen from solid and liquid 
surfaces, or from individual atoms. 
It has been noticed that the field emission from metals occurs in presence of high 
electric field: the gradients are typically higher than 1000 volts per micron 
and strongly dependent upon the work function. In modern science and 
technology, there are a lot of applications of field emissions, 
which includes construction of high-resolution and high luminosity 
electron microscopes and also used in modern nano science and
technology \cite{nano}.

Further, the thermal emission from metallic surface can be explained by
classical Boltzmann statistics, with only quantum correction needed is the
inclusion of Fermi energy. On the other hand, the photo-emission of electrons 
can not be explained unless the quantum nature of electromagnetic wave is
considered, i.e., old quantum mechanics or quantum theory is needed 
where photon concept of light is introduced. Of course, the concept of work function
associated with any kind of electron emission process is a quantum mechanical phenomenon. 
The field emission process on the other hand can only be explained by quantum 
tunneling of electrons. Which has no counter classical explanation. However, 
for general type of surface barrier, this purely quantum mechanical problem
can not be solved exactly, a semi-classical approximation, known as 
WKB is needed to get tunneling coefficients. Now because of quantum fluctuation,
electrons from the sea of conduction electrons (electron gas) always try to 
tunnel out through the metallic surface (surface barrier). However, as soon as an electron comes 
out, it induces an image charge on the metal surface, which pulls it back
and does not allow it to move far away in the atomic scale. But if a strong attractive
electrostatic field is applied near the metallic surface, then depending
on the Fermi energy of electrons and the height of the surface barrier, it may overcome the
effect of image charge and the electrons get liberated. Since the external
strong electric field is causing such emission and does not depend on
the thermal properties of the metal, even the metal can be at zero
temperature, it is called field emission or cold emission.

The theory of field emission from bulk metals was first proposed by
Fowler and Nordheim in an epoch making paper in the proceedings of
Royal Society of London in the year 1928 \cite{FN1,FN2,FN3}. 
Fowler-Nordheim tunneling is the wave-mechanical tunneling of electrons
through a triangular type barrier produced at the surface of an
electron conductor by applying a very high electric field. Fowler
Nordheim tunneling current is found to have important implication in
various scientific and technological applications, starting from cold
cathode emission to VLSI.  

Cold emission or field emission processes not only have significance in
the terrestrial laboratories, but is found to be equally important in the
electron emission processes from cold and compact stellar objects. 
For an example, in the case of a rotating neutron star, 
the existing large magnetic field, $\geq 10^{12}$G for the conventional
radio pulsars and $\geq 10^{15}$G for the magnetars,
produces a strong electric field 
at the poles, approximately given by $F\sim 2\times
10^8P^{-1}B_{12}$volt cm$^{-1}$, parallel to $B$ at the poles \cite{ST},
here $P$ is the time period of the neutron star in second. 
At the proximity of polar region the potential difference changes almost
linearly with distance from the polar cap, which is a region very
close to the magnetic pole. The repulsive surface barrier 
in combination with this attractive potential, forms a triangular type 
barrier at the poles. Now,
the study of plasma formation in a pulsar magnetosphere is a quite old 
but still an unresolved astrophysical issue. In the formation of 
magneto-spheric 
plasma, it is generally assumed that there must be an initial high energy 
electron flux from the magnetized neutron stars. At the poles of a neutron 
star the emitted charged particles flow only along the magnetic field lines.
The flow of high energy electrons along the direction of magnetic field
lines and their penetration through the light cylinder is pictured with the 
current carrying conductors. Naturally, if the conductor is broken near the 
pulsar surface the entire potential difference will be developed across a thin 
gap, called polar gap. This is of course based on the assumption that
above a critical height from the polar gap, because of high electrical 
conductivity of the plasma, the electric field $F$, parallel with the 
magnetic field near the poles is quenched. Further, a steady acceleration of
electrons 
originating at the polar region of neutron stars, traveling along the field 
lines, 
will produce magnetically convertible curvature $\gamma$-rays.
If these curvature $\gamma$-ray photons have energies $>2mc^2$ (with $m$ 
the electron rest mass and $c$ the velocity of light), then pairs of $e^--e^+$ 
will be produced in enormous amount with very high efficiency near the polar 
gap. These produced $e^--e^+$ pairs form what is known as the magneto-spheric 
plasma \cite{R2,R3,R4,R5,R6,R7,R8}. The cold emission, therefore plays a
significant role in magneto-spheric plasma formation. In turn, the
motion of charged particles in the magnetosphere in presence of strong magnetic field causes pulsar
emission in the form of synchrotron radiation. Therefore the cold emission process
indirectly affects the intensity of synchrotron radiation.

Although the field electron emission is a paradigm example of what
we call the quantum mechanical tunneling, unfortunately, it is also a paradigm
example of the intense mathematical difficulties associated with this
tunneling process with general type barrier structure. Exactly
solvable models with simple type tunneling barrier lead to equations
\cite{FN1,FN2} that underestimates the emission current density by a factor 
of 1000 or more. If a more realistic type barrier model is used by inserting 
an exact potential barrier in the simplest form of the 
Schr${\ddot{\rm{o}}}$dinger equation, then a complicated mathematical 
problem with very little physical important arises over the resulting 
differential equation: 
it is known to be mathematically impossible in principle to solve this 
equation exactly in terms of the usual functions of mathematical physics, 
or in any simple way. To get even an approximate solution, 
it is necessary to use special approximate methods known in physics as 
semi-classical or  WKB approximation. 

Moreover, to the best of our knowledge, a relativistic version of cold emission model in presence of 
strong magnetic field, relevant for electron emission
from the polar region of strongly magnetized neutron stars, even with a
simple type potential barrier has not been reported in the literature.
It has already been mentioned that 
the emission of high energy electrons and thereby formation of magnetosphere 
is caused by strong electric field at the polar region and
this strong electric field at the poles is produced by the magnetic field 
of rotating neutron stars.

We have organized the article in the following manner: in section-2 we have studied electron emission
with a scalar type potential barrier. In section-3 we have studied the same thing replacing the scalar
potential by a vector type potential barrier. In the last section we have given the conclusion and
future prospect of this work.
\section{Relativistic Emission Model in Presence of Magnetic Field:
Scalar Potential}
In the original paper by Fowler and Nordheim on cold emission, the conventional 
non-relativistic quantum mechanical equations, i.e., the Schr$\ddot{\rm{o}}$dinger 
equation was used \cite{FN1}. 
In the relativistic version we shall use the Dirac equation and introduce the 
triangular form of barrier potential at the surface of the
system using the conventional form of relativistic quantum hadro-dynamics. In the simplified picture
of quantum hadro-dynamic model the interaction can either be of scalar
type or vector type or both. In this model the conventional form of 
Dirac equation with scalar and vector type interactions is given by \cite{NKG}
\begin{equation}
\left [\gamma^\mu\left (i\partial_\mu-g_vV_\mu \right) -\left
(m-g_s\phi\right)\right]\psi = 0
\end{equation}
where $g_v$ and $g_s$ are the vector and scalar coupling constants respectively, 
$V_\mu$ and $\phi$
are respectively the vector and the scalar fields and $\gamma_\mu$s are
the Dirac $\gamma$-matrices. In presence of strong neutron star crustal 
magnetic field, electrons can travel almost along the field direction. If the magnetic field strength
$B\geq B_c^{(e)}$, where $B_c^{(e)}$ is the typical strength of magnetic field at which the electron
cyclotron quantum exceeds the corresponding rest mass energy, then in the relativistic scenario the
Landau levels of the electrons will be populated. In this situation, the electron momentum in the plane
orthogonal to the  direction of magnetic field, gets quantized and is given by $p_\perp=(2\nu
eB)^{1/2}$, where $\nu=0,1,2....$ are the Landau quantum numbers. Assuming that the magnetic field is
along $z$-direction the corresponding electron momentum component changes continuously from $-\infty$
to $+\infty$. It can be shown that the critical strength $B_c^{(e)}\approx 4.43\times 10^{13}$G.
In this situation the Dirac equation reduces to effectively
one-dimensional in nature. Further, we replace the scalar and 
vector part of the fields by the non-relativistic potentials at the
interface between matter and vacuum with triangular in shapes. 
In the case of vector potential, only the zeroth part is retained whereas 
its spacial part vanishes because of rotational symmetry within the
matter. In one-dimension, for the sake of
convenience we use the form of Dirac equation originally proposed by
Dirac, where $\gamma$-matrices 
are replaced by $\alpha$ and $\beta$ matrices. The triangular shape potential 
barrier is represented by the form $V(x)=C-eFx$, 
where following the notation of Fowler and Nordheim, $C$ is the surface barrier and
$F$ (absorbing the electron charge, we express $eF$ by $F$) is the
electric field at the polar region. The quantity $F$ acts as the driving force 
for electron emission. Let us first assume that the potential barrier is 
of scalar type. Then the Dirac equation in one-dimension 
is given by 
\begin{equation}
\left [\alpha p+\beta(m+C-Fx)\right]\psi=E\psi
\end{equation}
where $E$ is the energy eigen value. Following the representation \cite{R9}
\begin{equation}
\alpha=\sigma_y=\left( \begin{array}{l r}0 &-i\\ i& 0\\ \end{array} \right)
~~{\rm{and}}~~ \beta=\sigma_x=\left( \begin{array}{l r}0 &1\\ 1& 0\\ \end{array} \right)
\end{equation}
and defining $m^*=m+C$ as the effective electron mass, then with the spinor notation 
\begin{eqnarray}
\psi&=&\left (\begin{array}{c}u\\ v\\ \end{array}\right)\\ & {\rm{we~~have}} \nonumber \\
-\frac{dv}{dx}+(m^*-Fx)v&=&Eu\\
\frac{du}{dx}+(m^*-Fx)u&=&Ev
\end{eqnarray}
After eliminating $v$ from the eqns.(5)-(6), we have
\begin{equation}
-\frac{d^2u}{dx^2}+Fu+(m^*-Fx)^2 u=E^2 u
\end{equation}
To solve this equation analytically, we make the following transformation of coordinate:
\begin{equation}
\xi=F^{1/2}\left (\frac{m^*}{F}-x\right)
\end{equation}
Then we have 
\begin{equation}
-\frac{d^2 u}{d\xi^2}+\xi ^2u=\left (\frac{E^2}{F}-1\right)u
\end{equation}
This equation may be re-arranged in the following form 
\begin{equation}
\frac{d^2u}{d\xi^2}+(\lambda-\xi^2)u=0
\end{equation}
which is the well known Hermite differential equation, with
$\lambda=1+2\nu$, 
which gives $E^2=2(\nu+1)F$ and the solution is given by
\begin{equation}
u=N\frac{E}{F^{1/2}}\exp\left (-\frac{\xi^2}{2}\right) H_\nu(\xi)=u_I~~{\rm{(say)}}
\end{equation}
where $N$ is the normalization constant and $H_\nu(\xi)$ is the well known 
Hermite polynomial of order $\nu$ (which is not necessarily integer). 
It is trivial to show that $v$ will also have the same kind of solution. 
This spinor solution ($u$ or $v$) is for the electron which has already
been liberated through the surface into vacuum under the influence of 
electric field $F$.

Now the one-dimensional form of free Dirac equation for the electrons bound 
within the metal body (in this case within the metallic crustal matter), 
i.e., before emission from the system is given by 
\begin{equation}
\left (\alpha p+\beta m\right)\psi=E\psi
\end{equation}
Then following the same technique as above, we have 
\begin{equation}
\frac{d^2u}{dx^2}+w_k^2 u=0
\end{equation}
where $w_k=(E^2-m^2)^{1/2}$ (exactly same type of equation can be
obtained for $v$). Now following the notation of Fowler and Nordheim, 
we express the solution for free electrons within the system, 
confined by the surface barrier $V(x)$, in the form
\begin{equation}
u=\frac{1}{w_k^{1/2}}\left [a\exp(iw_kx)+a^\prime\exp(-iw_kx)\right]=
u_{II}~~{\rm{(say)}}
\end{equation}
where $a$ is the probability amplitude for electrons moving along the 
positive direction of $x$,
whereas $a^\prime$ is the corresponding quantity for left moving
waves. Assuming the interface between the neutron star matter
(which in this case is the dense metallic iron crystal of neutron star crustal matter) and vacuum is 
at $x=0$, the wave
function and their derivatives must be continuous at $x=0$, i.e., 
\begin{equation}
u_{II}(0)=u_I(0)~~~{\rm{and}}~~~ u_{II}^\prime(0)=u_I^\prime(0)
\end{equation}
Using the relation $H_\nu^\prime(\xi)=2\nu H_{\nu-1}(\xi)$, we have 
\begin{equation}
a+a^\prime=N w_k^{1/2}\frac{E}{F^{1/2}}\exp\left
(-\frac{\xi_0^2}{2}\right)H_\nu(\xi_0)
\end{equation}
and
\begin{equation}
iw_k^{1/2}(a-a^\prime)=NE\exp\left (-\frac{\xi_0^2}{2}\right)\left [\xi_0
H_\nu(\xi_0)-2\nu H_{\nu-1}(\xi_0)\right]
\end{equation}
where $\xi_0=\xi(x=0)=m^*/F^{1/2}$. These two conditions (eqns.(16)-(17)) 
may be rearranged in the following form
\begin{equation}
a+a^\prime=X ~~{\rm{and}}~~ a-a^\prime=iY
\end{equation}
where $X$ and $Y$ are two real quantities. Hence it is straight forward to
verify that the transmission coefficient, given by  
\begin{equation}
T_R=1-\frac{\vert a^\prime\vert^2}{\vert a\vert^2},
\end{equation}
vanishes exactly. In this section we therefore conclude that if the barrier 
in combination with the external electro-static driving force behaves like a 
scalar potential, electrons can not tunnel through the
surface barrier whatever be their kinetic energies and the strength of external electric field. Identical result will be 
obtained if $v$ is used instead of $u$. This result has some analogy with the
confinement of quarks inside hadronic bag with scalar type linear potential. 
\section{Relativistic Emission Model in Presence of Magnetic Field:
Vector Potential}
In presence of a vector type potential barrier at the interface with the same kind of triangular 
shape as mentioned in the previous section,
we have the conventional form of Dirac equation
\begin{equation}
(\alpha p+\beta m)\psi=(E-C+Fx)\psi
\end{equation}
Decomposing the spinor $\psi$ into $u$ and $v$ as before and using the same representation
for $\alpha$ and $\beta$, we have 
\begin{eqnarray}
\frac{du}{dx}+mu&=&(E-C+Fx)v\\
-\frac{dv}{dx}+mv&=&(E-C+Fx)u
\end{eqnarray}
Eliminating $v$ from these two equations (eqns.(21)-(22)) we have 
\begin{equation}
-\frac{d^2u}{dx^2}+\left
[\frac{du/dx+mu}{E-C+Fx}\right]+m^2u=(E-C+Fx)^2u
\end{equation}
Substituting $\xi=-(C-E)/F+x$, we have
\begin{equation}
\xi^2\frac{d^2u}{d\xi^2}-\xi \frac{du}{d\xi}-(m\xi+m^2\xi^2-F^2\xi^4)u=0
\end{equation}
To obtain an analytical solution for $u$, let us consider the non-linear transformation $\eta=\xi^2$.
Then the above equation becomes
\begin{equation}
\frac{d^2u}{d\eta^2}+u-\frac{1}{4\eta^2} \left (m\left
(\frac{2}{F}\right)^{1/2}\eta^{1/2}+\frac{2\eta m^2}{F}\right)u=0
\end{equation}
To obtain a closed analytical solution, we neglect for the sake of
simplicity the $\eta^{1/2}$ term. Then we have 
\begin{equation}
\frac{d^2u}{d\eta^2}+\left (1-\frac{2\chi}{\eta}\right)u=0
\end{equation}
where $\chi=m^2/4F$ and $\eta>0$. This is the standard differential equation for the
well-known Coulomb function with zero angular momentum \cite{abst}. 
The solution is given by 
\begin{equation}
u_0=\frac{C_0(\chi)}{2\eta}\exp(2i\delta_0)\exp[i(\eta-\chi \ln 2\eta)]=u_I
~~{\rm{(say)}}
\end{equation}
where $\delta_0={\rm{arg}}\Gamma(1+i\chi)$.\par
\bigskip\medskip
\hrule \par
\bigskip
{\it{The exact solution is quite complicated. A semi-compact solution may be 
obtained using Mapel package in terms of
HeunB function- the Heun Biconfluent function. This
function is related to Kummer function or Hypergeometric
function. The solution is given by}}:
\begin{eqnarray}
u(\eta)&=& C_1\exp(i\eta){\rm{HeunB}}(2,0,-2i\chi,(4+4i)\rho,(1-i)\eta^{1/2})
\eta +
C_2\exp(i\eta){\rm{HeunB}}(2,0,-2i\chi,(4+4i)\rho,(1-i)\eta^{1/2})\eta\nonumber \\ 
&&\times \left (\int \frac {\exp(-2i\eta)}{({\rm{HeunB}}(2,0,-2i\chi,(4+4i)\rho,(1-i)
\eta^{1/2}))^2 \eta^2}d\eta \right),\nonumber
\end{eqnarray}
{\it{where}} 
\[
\rho=\frac{m}{4}\left (\frac{2}{F} \right )^{1/2}
\]
\bigskip
\hrule \par
\bigskip\medskip
To obtain tunneling
co-efficient, we again consider the continuity of $u$ and its
derivative at the interface (i.e., at $x=0$). Which means 
\begin{equation}
u_{II}(0)=u_I(0)~~~{\rm{and}}~~~ u_{II}^\prime(0)=u_I^\prime(0)
\end{equation}
Hence we get 
\begin{equation}
(a+a^\prime)=w_k^{1/2}\frac{iX_0}{2\eta_0}
\end{equation}
where 
\begin{equation}
X_0=C_0\exp(2i\delta_0)\exp[i(\eta_0-\chi \ln 2\eta_0)]
\end{equation}
is the phase factor, and 
\begin{equation}
(a-a^\prime)=\frac{X_0}{\eta_0^{1/2}w_k^{1/2}}i\left
[(1-\frac{\chi}{\eta_0})+i\frac{1}{\eta_0}\right]
\end{equation}
Where, as before $\eta_0=\eta(x=0)$. Hence, from the definition of
tunneling coefficient, we have
\begin{eqnarray}
T_R&=&1-\frac{\vert a^\prime\vert^2}{\vert a\vert^2}\\
&=&\frac{2\omega_k\eta_0^{3/2}\left
(1-\frac{\chi}{\eta_0}\right)}{1+\eta_0\left
[\frac{\omega_k}{2}+\eta_0^{1/2}\left
(1-\frac{\chi}{\eta_0}\right)\right]^2}
\end{eqnarray}
Substituting $\eta_0=(C-E)^2/2F$, we finally get 
\begin{equation}
T_R=\frac{2^{5/2}F^{1/2}\omega_k(C-E)^3\left
(1-\frac{m^2}{2(C-E)^2}\right)}{8F^2+(C-E)^2\left [\omega_k
F^{1/2}+2^{1/2}(C-E)\left (1-\frac{m^2}{2(C-E)^2}\right)\right]^2}
\end{equation}
Obviously, in this case, unlike the scalar type surface barrier, 
the transmission co-efficient is non-zero. 
Further, unlike Fowler Nordheim non-relativistic model, here the tunneling 
coefficient dose not fall exponentially. In the above expression for
tunneling coefficient, as before  
$C=\mu+W_f$ and $\omega_k=(E^2-m_\nu^2)^{1/2}$, with $m_\nu=(m^2+2\nu
eB)^{1/2}$ and $\nu=0,1,2........$ are the Landau quantum numbers for the
electrons. In presence of magnetic field, it has been observed that the work function increases 
with the increasing field strength. A numerically fitted form for
magnetic field dependence 
is given by $W_f=W_c (B/B_c^{(e)})^{0.5}$ in eV, where $W_c$ is a real constant parameter 
$\approx 82.93$ \cite{lai,epjd}. 
One can easily verify that the tunneling coefficient will vanish
for $F=0$ and extremely large values for electron work function.
\section{Relativistic Form of Cold Emission Current for $B\neq 0$}
To obtain electron cold emission current in the relativistic scenario in
presence of strong magnetic field, we consider the motion of
relativistic electrons along the direction of magnetic field and the
system is assumed to be at zero temperature.
It is well known that in presence of strong quantizing
magnetic field $\geq 10^{15}$G, which is the recently observed surface 
values of some strange type neutron stars, known as magnetars \cite{mag}, 
the motion of the electrons at the crustal region of a neutron star is 
effectively one dimensional 
in nature; motion is restricted along the direction of flux lines. In
this case the density of states is given by \cite{HW}
\begin{equation}
\frac{d^3p}{(2\pi)^3}=\frac{eB}{4\pi^2}\sum_{\nu=0}^\infty(2-\delta_{\nu 
0})dp_z
\end{equation}
In this expression we have taken Planck constant $\hbar=1$, velocity of light 
$c=1$ and Boltzmann
constant $k_B=1$; known as natural unit. Here $\nu$ is the 
Landau quantum number and the factor $2-\delta_{\nu 0}$ takes care of 
singly degenerate zeroth Landau level and doubly degenerate other levels 
with $\nu \neq 0$. The upper limit of Landau quantum number is infinity 
at finite temperature, whereas at zero temperature it is given by 
\begin{equation}
\nu_{max}=\left [\frac{\mu^2-m^2}{2eB}\right]
\end{equation}
here $\mu$ is the electron chemical potential and $[~]$ indicates that
$\nu_{max}$ is an integer but less than the actual number. Now the number of
electrons incident normally on unit area at the interface
between matter and vacuum is given by 
\begin{equation}
R=\frac{eB}{4\pi^2}\sum_{\nu=0}^\infty(2-\delta_{\nu 0})\int_{p_0}^\infty
\frac{p_z}{E_\nu}\frac{dp_z}{\exp\left [\frac{E_\nu-\mu}{T}\right]+1}
\end{equation}
where $E_\nu=(p_z^2+m^2+2\nu eB)^{1/2}$ is the electron energy at
$\nu$th Landau level and $p_0$ is the lower limit of $z$ component of 
electron momentum.
Unlike thermionic emission and photo-emission, the cold emission is a
purely quantum mechanical phenomenon; electrons tunnel through the
surface barrier and its emission in free space is driven by the external
electric field. Since the emission process is controlled by the tunneling
probability of electrons through surface barrier, the actual 
expression for cold emission current will therefore be 
\begin{equation}
R=\frac{e^2B}{\pi^2}\sum_{\nu=0}^\infty(2-\delta_{\nu_0})\int_0^\infty\frac{p_z}{E_\nu}
\frac{dp_z}{\exp\left [\frac{E_\nu-\mu}{T}\right]+1}\times T_R
\end{equation}
In the case of neutron star, since the temperature is low enough compared to the
chemical potential of electrons, we put $T=0$ in the Fermi distribution in eqn.(38), then we
have after replacing the integration variables from $p_z$ to $E_\nu$ and omitting the index
$\nu$ (keeping in mind that $E=E(p_z,\nu,B)$) in electron energy 
\begin{equation}
R=\frac{e^2B}{4\pi^2}\sum_{\nu=0}^{\nu_{max}}(2-\delta_{\nu 0})\int_{m_\nu}^\mu T_R dE
\end{equation}
We have obtained cold emission current by evaluating numerically energy integral and
the sum over Landau quantum number. However, we have noticed that at the crust of magnetars the Landau
quantum number becomes exactly zero, i.e., the electrons occupy only their zeroth Landau level. In fig.(1) we have shown the variation of cold
current with electric field and magnetic field by the solid curve and the dashed curve
respectively. Both the curves indicate the initial increase of cold current. Since the
work function increases with increasing magnetic field and electric field 
is generated by the rotating magnetic field of neutron stars, the cold current decreases 
abruptly after some magnetic field / electric field strength.
\section{Conclusions} 
In this article we have studied cold emission of electrons from strongly magnetized neutron star polar
region using relativistic quantum mechanics. To get an analytical solution, we have considered triangular
type potential barrier as was used by Fowler and Nordheim. We have noticed that a scalar type
potential barrier does not allow electrons to tunnel out the interface, whatever be the strength of
external electric field. With the vector type potential barrier, we have obtained an analytical
solution for the tunneling coefficient. Unlike the original work of Fowler and Nordheim the tunneling
coefficient does not follow any exponential law. However, the variation
of cold current with the electric field strength can be fitted
numerically using a $\chi^2$-minimization program. The numerically
fitted functional form is given by 
\begin{equation}
R=0.26 F_{24}^{1/2} \exp(-9.8 \times F_{14})
\end{equation}
where $F_{14}=10^{-14}F$ and $F_{24}=10^{-24}F$. The variation shows an
exponential decay form is because of approximate linear dependence of
electric field with the magnetic field strength of the rotatiting neutron
star.
We have obtained an analytical expression for 
the cold emission current which changes with the strength of magnetic field also. We found that the
cold current initially increases almost linearly with the external electric field and finally when the
work function becomes large enough because of high magnetic field strength, it drops abruptly and
finally goes to zero. Since the emission current becomes extremely small in presence of ultra-strong
magnetic field, the magnetosphere of the magnetars will be of very low in charge density. As a
consequence there will be very weak synchrotron emission in the radio wave band.

\begin{figure}
\psfig{figure=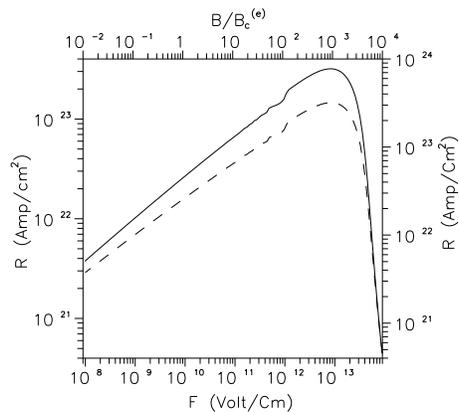,height=0.3\linewidth}
\caption{Variation of cold emission current with neutron star magnetic field (dashed
curve) expressed in terms of critical field $B_c^{(e)}$ for electron and electric
field in Volt/cm (solid curve) at the neutron star pole.}
\end{figure}

\begin{thebibliography}{99}
\bibitem{nano} Shi-Dong Liang and Lu Chen, Phys. Rev. Lett., 101, 027602 (2008).
\bibitem{FN1} R.H. Fowler and Dr. L. Nordheim, Proceedings of the Royal Society 
of London 119, 173, 
(1928).
\bibitem{FN2} T.E. Stern, B.S. Gossling and R.H. Fowler, Proceedings of the 
Royal Society of London, 124, 699, (1929).
\bibitem{FN3} K. L. Jensen, J. Vac. Sci. Technol., B13, 516, (1995); see
also R. G. Forbes and J. H. B. Deane, Proceedings of the
Royal Society of London, 463, 2907, (2007).
\bibitem{ST} S.L. Shapiro and S.A. Teukolsky, Black Holes, White Dwarfs,
and Neutron Stars- The Physics of Compact Objects, John Wiley \& Sons,
New York, 288, (1983).
\bibitem{R2} A. Jessner, H. Lesch and T. Kunzl, APJ, 547, 959, (2001). 
\bibitem{R3} M.A. Ruderman and P.G. Sutherland, APJ, 196, 51, (1975).
\bibitem{R4} see also, D.A. Diver, A.A. da Costa, E.W. Laing, C.R. Stark and L.F.A. Teodoro,
astro-ph/0909.3581. 
\bibitem{R5} S.L. Shapiro and S.A. Teukolsky, Black Holes, White Dwarfs
and Neutron Stars, John Wiley and Sons, New York, (1983).
\bibitem{R6} F.C. Michel, Rev. Mod. Phys., 54, 1, (1982); F.C. Michel, Advances in Space Research,
33, 542, (2004).
\bibitem{R7} A.K. Harding and D. Lai, Rep. Prog. Phys., 69, 2631, (2006).
\bibitem{R8} M. Ruderman, Phys. Rev. Letts., 27, 1306, (1971).  
\bibitem{NKG} N. K. Glendenning, Compact Stars (Second Edition),
Springer, 162, (2000).
\bibitem{R9} R. Jackiw and C. Rebbi, Phys. Rev., D13, 3398, (1976) (see
also R. S. Bhalerao and B. Ram, Am. J. Phys., 69, 817, (2001)). 
\bibitem{abst} M. Abramowitz and I. A. Stegun, Handbook of Mathematical
Functions, Dover Publications, INC., New York, 538, (1970).
\bibitem{mag} S. Mereghetti, astro-ph/0904.4880.
\bibitem{lai} Z.Medin and D.Lai, MNRAS, 382, 1833, (2007); Z.Medin and D.Lai, Phys. Rev. A74, 062507, (2006);
A.Harding and D.Lai, astro-ph/0606674.
\bibitem{epjd} A. Ghosh and S. Chakrabarty (submitted).
\bibitem{HW} K. Huang, Statistical Mechanics, Wiley Estern Private Ltd., New Delhi, 237, (1963).
\bibitem{AS1} A. Ghosh and S. Chakrabarty (to be communicated).
\end{thebibliography}
\end{document}